\documentclass[%
 reprint,
 amsmath,amssymb,
 aps,
 prl,
]{revtex4-1}
\usepackage{float}
\usepackage{graphicx}% Include figure files
\usepackage{dcolumn}% Align table columns on decimal point
\usepackage{bm}% bold math
\begin{document}

\preprint{APS/123-QED}

\title{Ideal magnetic dipole scattering}% Force line breaks with \\

\author{Tianhua Feng$^{1}$, Yi Xu$^{1,2}$}
\email{e$_$chui@qq.com}
\author{Wei Zhang$^{1}$}
\author{Andrey E. Miroshnichenko$^{3}$}
\email{andrey.miroshnichenko@anu.edu.au}
\affiliation{$^{1}$JNU-IREE Joint Laboratory of Fractal Method and Signal Processing, Department of Electronic Engineering, College of Information Science and Technology, Jinan University, Guangzhou, 510632, China}
\affiliation{$^{2}$Key Laboratory of Optoelectronic Information and Sensing Technologies of Guangdong Higher Education Institutes, Guangzhou, Guangdong 510632, China}
\affiliation{$^{3}$Nonlinear Physics Centre, Research School of Science and Engineering, The Australian National University, Acton, ACT 2601, Australia}

\date{\today}% It is always \today, today,

\begin{abstract}
We introduce the concept of tunable \textit{ideal magnetic dipole scattering}, where a nonmagnetic nanoparticle scatters lights as a pure magnetic dipole. High refractive index subwavelength nanoparticles usually support both electric and magnetic dipole responses. Thus, to achieve ideal magnetic dipole scattering one has to suppress the electric dipole response. Such a possibility was recently demonstrated for the so-called anapole mode, which is associated with zero electric dipole scattering. By overlapping magnetic dipole resonance with the anapole mode we achieve ideal magnetic dipole scattering in the far-field with tunable high scattering resonances in near infrared spectrum. We demonstrate that such condition can be realized for two subwavelength geometries. One of them is core-shell nanosphere consisting of Au core and silicon shell. It can be also achieved in other geometries, including nanodisks, which are compatible with current nanofabrication technology. 
\begin{description}
\item[PACS numbers]
42.25.-p, 78.67.-n, 42.25.Fx
\end{description}
\end{abstract}

%\pacs{}% PACS, the Physics and Astronomy
\maketitle

%\tableofcontents

%\section{Introduction}electric dipole

Light-matter interaction mainly relies on the electric component of light, especially for nonmagnetic materials \cite{H1,H2,H3,H4,H5}. The responses of such materials to the magnetic component of light is usually negligible at frequencies above a few terahertz \cite{weak1,weak2,weak3}. Therefore, realization of effective optical magnetism is very crucial for designing functional metamaterials working at this spectrum range \cite{PRL2012,nt,PRL2015}. Optical Mie resonances in dielectric nanoparticles with high refractive indices provide a platform for studying magnetic light-matter interaction mediated by their optical induced magnetic resonances \cite{science,OE11,ML,Si_NL12}. Such a nanoparticle supports a magnetic dipole (MD) mode which resembles a circular displacement current distribution in the near-field and therefore manifests itself as a MD scatter in the far-field. To date, various promising functions have been demonstrated via involving MD modes, such as directional scattering of light \cite{shawei,wei,directional1,directional2,AE_nano,Belov,min,chen1,yang2}, Fano resonances \cite{Andrey_NL,small,yang1,core_shell_oe}, enhanced optical nonlinearity \cite{enhanced}, nonradiate anapole \cite{anapole}, subwavelength topological edge states \cite{topological}, Purcell factor enhancements \cite{Purcell1,Purcell2,Purcell3,Purcell4}, and metasurface \cite{metasurface1,metasurface2} etc.

In general, the resonant frequencies of electric dipole (ED) and MD modes in silicon nanoparticles (SNs) are shifted with respect to each other. Thus, they spectraly overlap in quite significant region, making SN as a 'dressed' MD scatter \cite{ML}. As a result, they can exhibit various interference effects in the far-field, leading to the unidirectional Fano resonant scattering of light \cite{wei,directional1,directional2}. Therefore, in order to unambiguously study the MD scattering, it is necessary to suppress the electric dipole and all higher-order multipoles. Such kind of exotic ideal MD scatterer would be extremely useful for the study of \textit{magnetic light-matter interaction}. 

Regarding a SN, the conditions to realize an ideal MD scatter are the solutions of the following equation set (details can be found in the Supplemental Material \cite{Si}): 

\begin{eqnarray}
\label{eq:eq1}
\left\lbrace
\begin{aligned}
|(D_{1}(mkR)/m+1/(kR))\psi_{1}(kR)-\psi_{0}(kR)|\ll 1\\
|b_{1}(\lambda)|_{R}'=0\;,\;\;|b_{1}(\lambda)|_{R}''<0\\
|a_{n}(\lambda,R)|\;,\;\;|b_{n}(\lambda,R)|\ll 1,n>1
\end{aligned}
\right.
\end{eqnarray}
where $k$ = 2$\pi/\lambda$ is the free space wavenumber, $\lambda$ is the wavelength in vacuum, $R$ is the radius of the SN, $m$ is the relative refractive indices of Si with respect to the ambient medium, $D_{n}(mz)=\psi_{n}'(mz)/\psi_{n}(mz)$,$\psi_{n}(z)$ is the Riccati-Bessel function and $b_{1}$ is the Mie scattering coefficient. Thus, we have just two free parameters - incident wavelength $\lambda$ and radius $R$. However, one can easily check that this set of equations does not have a solution for $\lambda$, and $R$ in the visible and near infrared wavelength range. It can be understood due to the fact that these two parameters are related to each other via so-called size parameter $q=2\pi R/\lambda$ and by simply changing the radius of SN will simultaneously shift the resonances of the ED and MD along the same spectrum direction and the spectrum overlap between ED and MD is always preserved \cite{ML,Si_NL12}. But, in general, there is a possibility to achieve an ideal MD scattering via overlap the MD mode with the nonradiated anapole mode which had been demonstrated very recently in SNs \cite{anapole}. It has also been proved that such anapole can be effectively tuned in Ag core/silicon shell nanoparticle to realize scattering transparency \cite{wei_lpr}. Therefore, it is expected that core/shell nanoparticles could provide the possibility to suppress the ED mode at the MD resonance.

In this Letter, we introduce the concept of \textit{ideal MD scatter scattering} whose ED and high order electric and magnetic multipoles are negligible at the resonant frequency of MD. Such suppression is based on the excitation of the anapole mode, which is a result of the destructive interference between Cartesian ED and toroidal dipole (TD) modes \cite{anapole,wei_lpr}, in an Au core/silicon shell nanoparticle. We show that there are two ED modes in this hybrid system and their interaction can be used to shift the resonant frequency of the anapole mode to overlap with the MD mode. We demonstrate both analytically and numerically that the total scattering of an Au core/silicon shell nanoparticle is completely dominated by its MD mode which manifests itself as an ideal MD scatter based on nonmagnetic materials. Such system can facilitate the study of magnetic light-matter interaction in nanoscale.

\begin{figure}[h]
\includegraphics[width=\columnwidth]{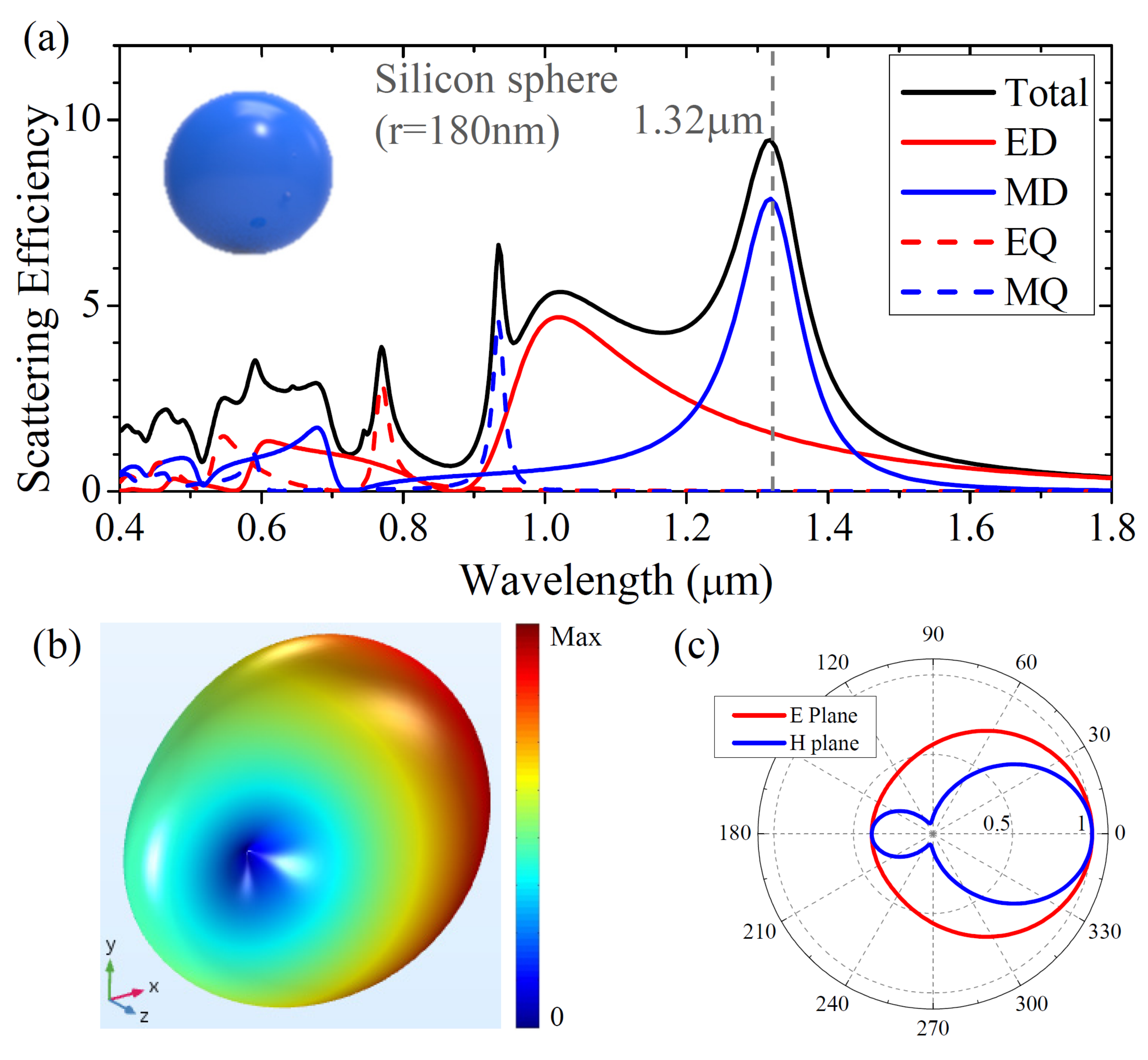}
\caption{\label{fig:fig1} 
(Color online) (a) Scattering efficiency of a silicon sphere embedded in air with a radius $R$ of 180 nm. The contributions of spherical ED, MD, electric quadrupole (EQ) and magnetic quadrupole (MQ) to the total scattering are presented. (b) Far-field scattering pattern at the MD resonant wavelength marked by a dashed line in (a). A plane wave is impinged along x-axis with its electric polarization along y-axis. (c) Cross-section views on the x-y plane (E plane) and x-z plane (H plane) of the far-field scattering pattern shown in (b).
}
\end{figure}

%\section{Pure magnetic dipole scattering}

As mentioned above, the spectral overlap of ED and MD modes is always present. In Fig. 1 (a) we show  the results based on the Mie theory \cite{book} for a SN of radius $ R $ = 180 nm. The MD resonance occurs when the size of SN satisfies the condition $ \lambda_{0}/n_{Si}\approx 2R_{Si} $ \cite{OE11,ML,Si_NL12}, where $ R_{Si} $ is the radius of the SN, $ n_{Si} $ is the refractive index of silicon and $ \lambda_{0} $ is the free space wavelength. Considerable spectrum overlap between ED and MD modes in the range 900-1600 nm is observed. It can be seen from Figs. 1 (b) and 1(c) that the far-field scattering pattern at the resonant frequency of MD is asymmetric and deviates from an ideal dipole type scattering pattern with non-zero scattering intensities along magnetic dipole axis at $ 90^{o} $ and $ 270^{o} $. It means that the ED component contributes to the far-field scattering and the SN cannot be regarded as an ideal MD scatter. 

\begin{figure}[!htb]
\includegraphics[width=\columnwidth]{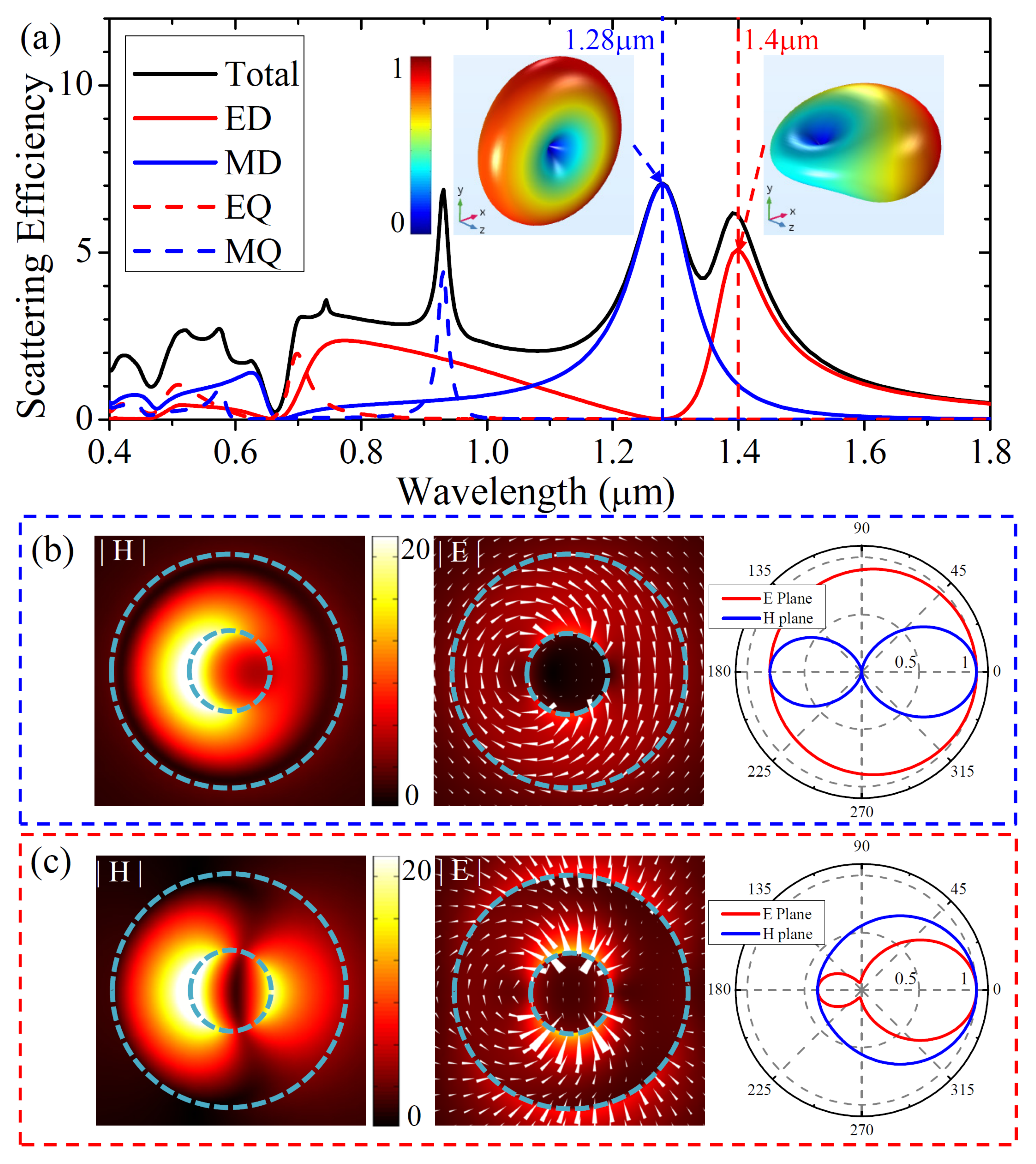}
\caption{\label{fig:fig2} 
(Color online) (a) Scattering efficiency of a Au(core)/silicon(shell) nanoparticle embedded in air with an outer radius $R$ = 180 nm and an inner radius $r$ = 62 nm. The contributions of spherical ED, MD, EQ and MQ to the total scattering are also presented. A plane wave is incident along x-axis with its electric polarization along y-axis. The far-field scattering patterns at the MD and ED resonant wavelengths indicated by dashed lines are presented in the insets. The induced electric $|E|$ and magnetic $|H|$ near-field distributions, which are obtained by subtracting the incident field from the total field, are shown for MD (b) and ED (c) modes. The corresponding cross section views of the far-field scattering patterns in (a) are also shown in (b) and (c).
}
\end{figure}

\begin{figure}[!htb]
\includegraphics[width=\columnwidth]{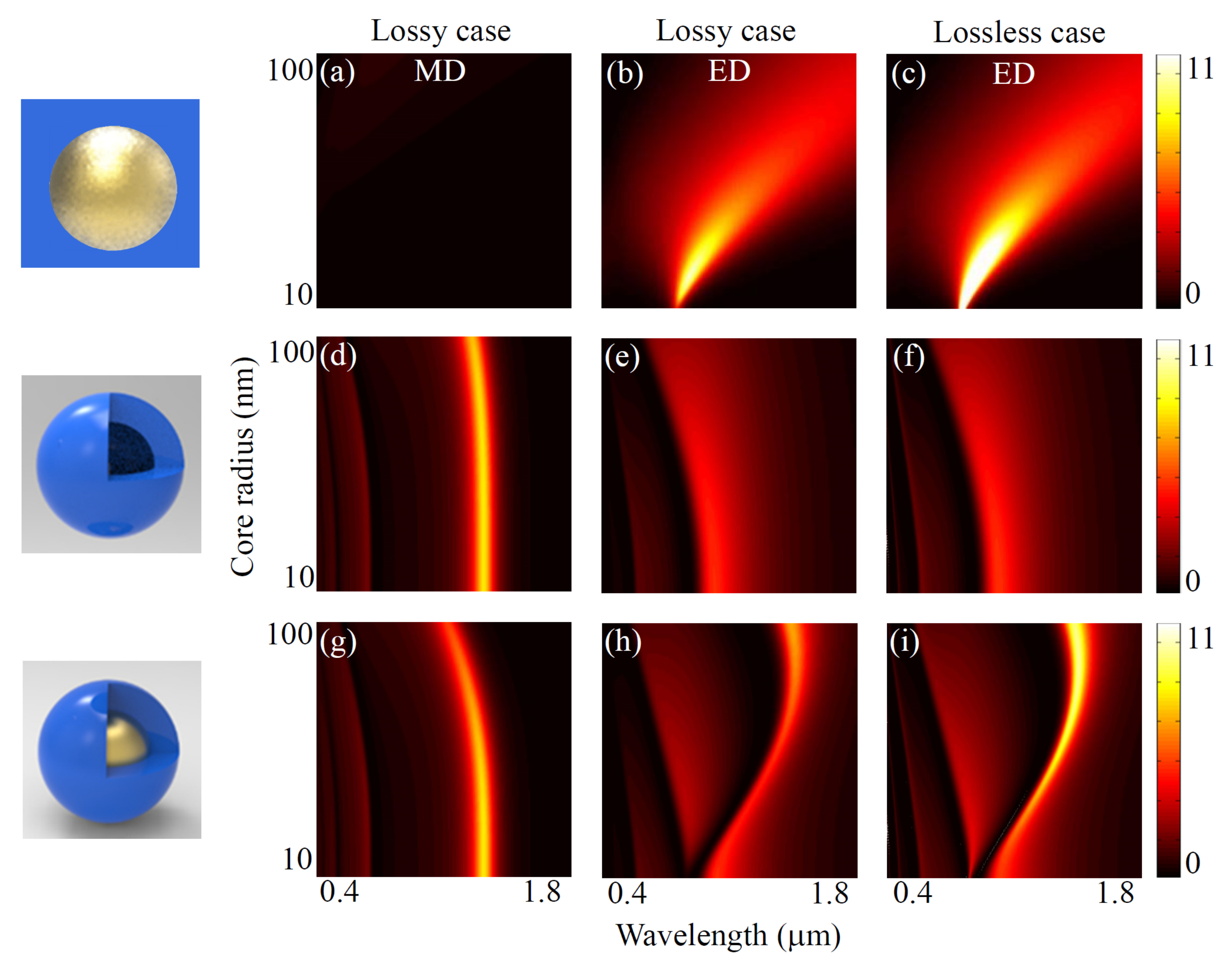}
\caption{\label{fig:fig2} 
(Color online) The first column: Sketch of various nanoparticle geometries, which show the Au nanosphere embedded in silicon, silicon shell with a void and Au(core)/silicon(shell) nanoparticle embedded in air. The outer radius $ R $ of silicon shells are 180 nm; the second column: the corresponding spherical MD contributions to the total scattering referring to the structure on the left; the third column: the corresponding spherical ED contributions to the total scattering for the lossy case; The fourth column: the corresponding spherical ED contributions to the total scattering for the lossless case.
}
\end{figure}

For a typical Au(core)/silicon(shell) nanoparticle, the degree of spectrum overlap between the ED and MD resonances can be reduced (see Fig. S1 in the Supplemental Material \cite{Si}). Compared with the SN of the same outer radius shown in Fig. 1, the MD resonant wavelength is almost unchanged and the MD contribution to total scattering is only slightly reduced as the Au core is located at the node of electric field for the MD mode. Most importantly, the anapole mode appears around $\sim\lambda $ = 1.25 $ \mu $m (see Fig. S1 in the Supplemental Material \cite{Si}). These results suggest that there is a possibility to overlap the anapole and MD modes in a core/shell nanoparticle. Analytically, the conditions to obtain \textit{ideal MD scattering} of a core/shell nanoparticle are the solutions of the following equation set:
\begin{eqnarray}
\label{eq:eq2}
\left\lbrace
\begin{aligned}
|(\widetilde{D}_{1}/m_{2}+1/(kR))\psi_{1}(kR)-\psi_{0}(kR)|\ll 1\\
|b_{1}(\lambda)|_{r,R}'=0\;,\;\;|b_{1}(\lambda)|_{r,R}''<0\\
|a_{n}(\lambda,r,R)|\;,\;\;|b_{n}(\lambda,r,R)|\ll 1,n>1
\end{aligned}
\right.
\end{eqnarray}
where 
\begin{eqnarray}
\label{eq:eq3}
\widetilde{D}_{1}=\frac{D_{1}(m_{2}kR)-A_{1}\chi_{1}'(m_{2}kR)/\psi_{1}(m_{2}kR)}{1-A_{1}\chi_{1}(m_{2}kR)/\psi_{1}(m_{2}kR)}
\end{eqnarray}
\begin{eqnarray}
\label{eq:eq4}
\widetilde{G}_{1}=\frac{D_{1}(m_{2}kR)-B_{1}\chi_{1}'(m_{2}kR)/\psi_{1}(m_{2}kR)}{1-B_{1}\chi_{1}(m_{2}kR)/\psi_{1}(m_{2}kR)}
\end{eqnarray}
\begin{eqnarray}
\label{eq:eq5}
A_{1}=\frac{\psi_{1}(m_{2}kr)(mD_{1}(m_{1}kr)-D_{1}(m_{2}kr))}{mD_{1}(m_{1}kr)\chi_{1}(m_{2}kr)-\chi_{1}'(m_{2}kr)}
\end{eqnarray}
\begin{eqnarray}
\label{eq:eq6}
B_{1}=\frac{\psi_{1}(m_{2}kr)(D_{1}(m_{1}kr)/m-D_{1}(m_{2}kr))}{D_{1}(m_{1}kr)\chi_{1}(m_{2}kr)/m-\chi_{1}'(m_{2}kr)}
\end{eqnarray}
where $k$ = 2$\pi/\lambda$ is the free space wavenumber, $\lambda$ is the wavelength in vacuum, $r$ and $R$ are the inner and outer radius of the core/shell nanoparticle and $m = m_{2}/m_{1}$ with $m_{1}$ and $m_{2}$ are the relative refractive indices of inner and outer materials relative to the ambient medium. 

One set of the typical solutions of Eq. (2) in the optical spectrum range, 600 nm $<$ $\lambda$ $<$ 1600 nm, are $\lambda$ = 1280 nm, $r$ = 62 nm and $R$ = 180 nm (see Fig. S2 in the Supplemental Material \cite{Si} for other solutions). Fig. 2 (a) presents the results for this case. We can see that the wavelength where the peak of the spherical MD contribution to the total scattering locates is exactly equal to the zero response of spherical ED. Meanwhile, the contributions from EQ and MQ modes are negligible. The far-field scattering results shown in Figs. 2(a) and 2(b) also validate a pure MD response, exhibiting as a doughnut shape and zero scattering intensities along the magnetic dipole axis ($ 90^{o} $ and $ 270^{o}$). For comparison, we also provide the far-field scattering results at the peak of the ED mode in Figs. 2(a) and 2(c). They obviously deviate from the ideal ED scattering because of the spectrum overlap between ED and MD modes.

In order to further explore the underlying physics for the overlapping of the anapole and MD modes in a core/shell nanoparticle, we inspect the scattering of an Au nanoparticle embedded in a silicon background first, as shown in Figs. 3(a)-(c). As expected, there is no MD response for the Au particle (see Fig. 3(a)). The dependence of the contribution of spherical ED to the total scattering on the size of the Au nanoparticle is shown in Fig. 3(b). One can see that there is a lower limit for the resonant wavelength of the ED mode ($\lambda$ = 600 nm). We further consider the case of a silicon shell ($ R $ = 180 nm) embedded in vacuum with a varied size of air core (void) in Figs. 3(d)-(f). As can be seen from these figures, ED and MD resonances are blue shifted together with the increasing of the void size due to the reduction of effective permittivity. Finally, we inspect the case of an Au core/silicon shell nanoparticle shown in Figs. 3(g)-(i). It is shown that there is a bifurcation point in the spherical ED decomposition results when we increase the radius $ r $ of Au core from 10 nm to 25 nm [see Figs. 3(h) and 3(i)], indicating strong interaction between two ED modes and the effective excitation of the anapole mode (destructive interference between Cartesian ED and TD modes) \cite{anapole}. Most importantly, we can conclude from Figs. 3(h) and 3(i) that the anapole mode can be fine-tuned from 600-1600 nm which can resonantly overlap with the MD mode at the near infrared spectrum shown in Fig. 3 (g).

\begin{figure}[!htb]
\includegraphics[width=0.8\columnwidth]{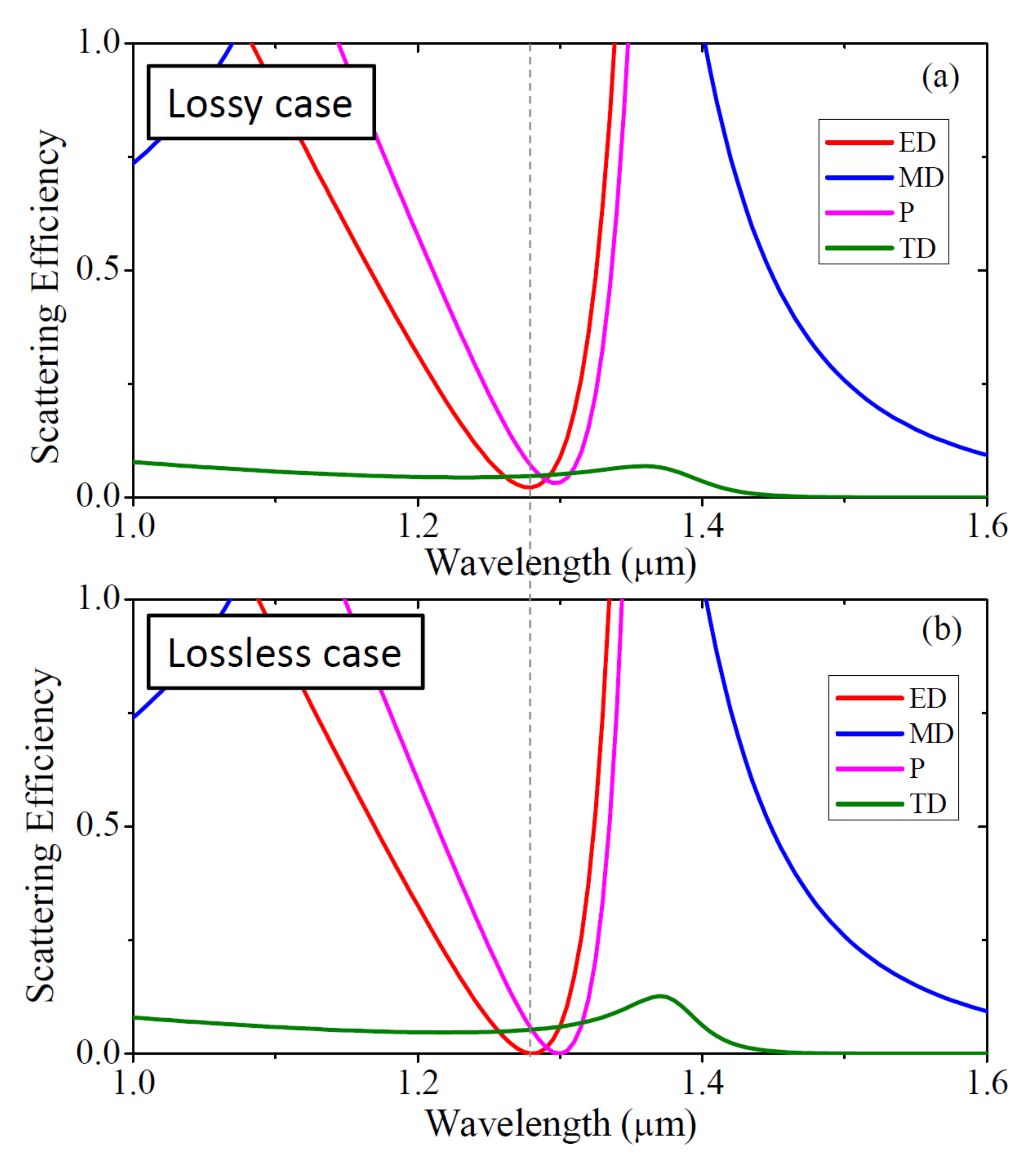}
\caption{\label{fig:fig4} 
(Color online) Spherical ED (red) and MD (blue), Cartesian ED (P, purple) and TD (green) moments contributions to the total scattering of the core/shell nanoparticle shown in Fig. 3 for lossy (a) and lossless (b) cases. Vertical dashed line indicates the resonant frequency where the ideal MD scattering appears.
}
\end{figure}

In order to validate the effective excitation of the anapole mode in this core/shell nanoparticle, we calculate the scattering efficiency of Cartesian ED and TD as \cite{TD1,TD2,TD3,TD4}

\begin{eqnarray}
\label{eq:eq7}
\sigma_{P}=\frac{\mu_{0}\omega^{4}}{12\pi^2 R^2 c}\vert \textbf{P}\vert^{2},\sigma_{T}=\frac{\mu_{0}\omega^{4}k^{2}}{12\pi^2R^2 c}\vert \textbf{T}\vert^{2}
\end{eqnarray}
where 
\begin{eqnarray}
\label{eq:eq8}
\textbf{P}=\frac{1}{-i\omega}=\int d^{3}r\textbf{J}(\textbf{r}),\\
\textbf{T}=\frac{1}{10c}=\int d^{3}r[(\textbf{r}\cdot \textbf{J}(\textbf{r}))\textbf{r}-2r^{2}\textbf{J}],
\end{eqnarray}
$ \mu_{0} $ is the permeability of vacuum, $ \omega $ is the angular frequency, $ c $ is the speed of light and $ \textbf{r} $ specifies the location where the induced current $ \textbf{J} $ is evaluated. As can be seen in Fig. 4, there is a cross point for the scattering efficiency of Cartesian ED and TD contributions, implying that the condition ($\textbf{P} = -ik\textbf{T}$) when the Cartesian ED and TD can cancel the scattering of each other has been fulfilled \cite{anapole,wei_lpr}. The slight difference between the wavelengths where the contribution of Cartesian ED is equal to TD and the minimum response of spherical ED is due to the material losses. Figure 4 (b) presents the lossless case in which the cross point of Cartesian ED and TD are exactly the same with the zero response of spherical ED, indicating the resonant excitation of the anapole mode at the resonance of the MD mode, i.e. \textit{ideal MD scattering}. It should be noted here that we only mimic the pure MD scattering in the far-field, since the near-field distribution of electromagnetic fields shown in Fig. 2 (b) is a superposition of MD and anapole modes. Since a planar nanostructure is more compatible with the state-of-the-art nanofabrication technologies for silicon photonics, a planar ideal magnetic scatter is of great demand compared with the spherical core/shell nanoparticle. We further provide a design for the ideal MD scattering based on a nanodisk. The plane wave scattering results of the Au(core)/silicon(shell) nanodisk shown in Fig. 5 confirm that such structure can also realize the \textit{ideal MD scattering} in the near infrared wavelength range (see Fig. S3 in the Supplemental Material \cite{Si} for the decomposition results). We also investigate the effect of imperfections on the ideal MD scattering. We model the imperfections in the fabrication process with 5 nm position variation of the Au core according to the state-of-the-art fabrication technology. The calculated results demonstrate that the proposed \textit{ideal MD scattering} is still robust when moderate fabrication imperfections are taken into account (see Fig. S4 in the Supplemental Material \cite{Si}).

\begin{figure}[!htb]
\includegraphics[width=\columnwidth]{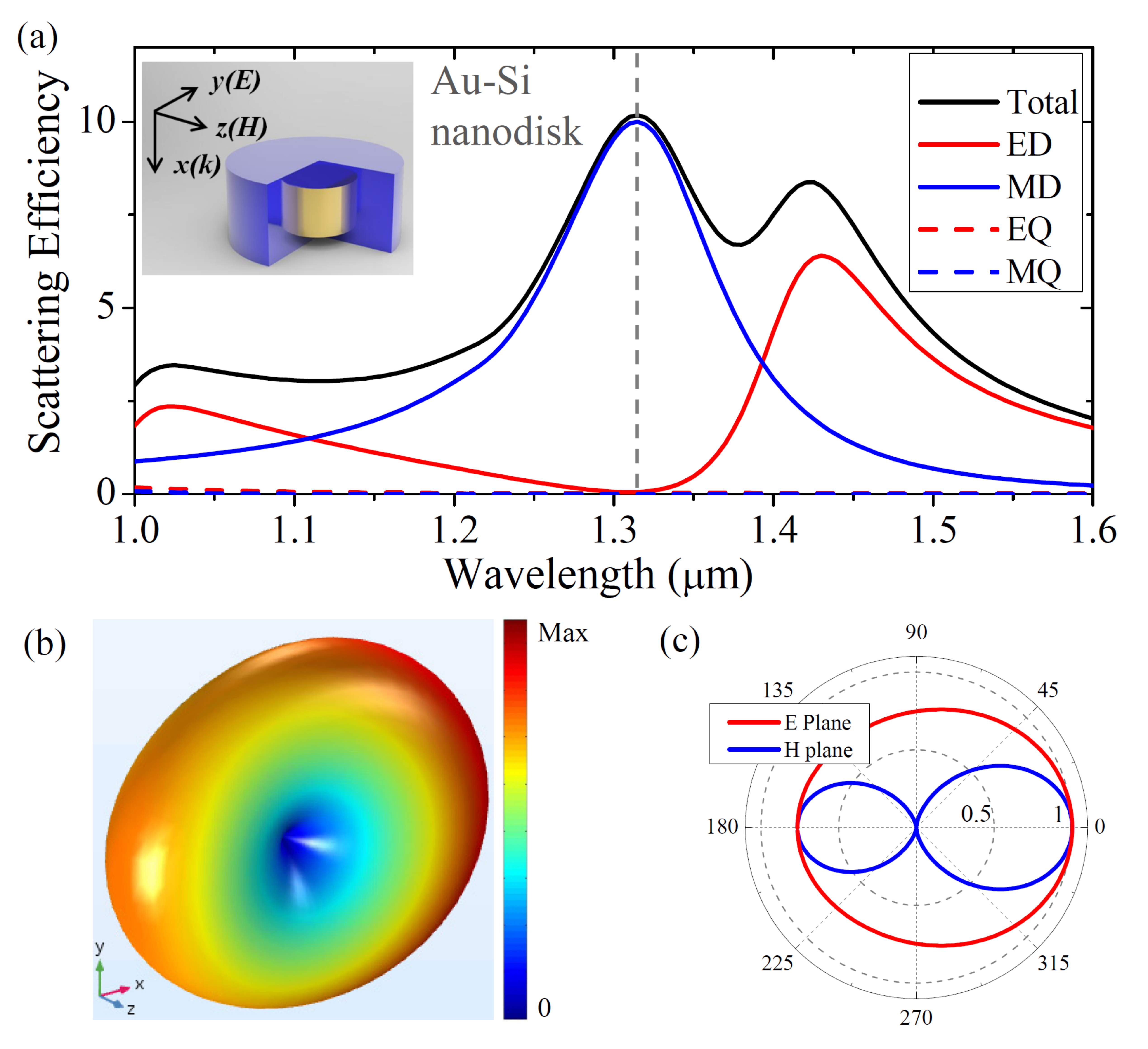}
\caption{\label{fig:fig5} 
(Color online) (a) Scattering efficiency of a Au(core)/silicon(shell) nanodisk embedded in air with the outer radius $R$ = 160 nm, outer height $H$ = 350 nm and an inner radius $r$ = 50 nm, inner height $h$ = 125 nm. The contributions of spherical ED, MD, EQ and MQ to the total scattering are also presented based on the decomposition utilizing spherical harmonics \cite{multipole}. Schematic shows the geometry of the nanodisk. A plane wave is incident along the axis of the cylinder (x-axis) with its electric polarization along y-axis. (b) The far-field scattering pattern at the ideal MD scattering wavelength indicated by the dashed line in (a). (c) Cross section views on the x-y plane (E plane) and x-z plane (H plane) of the far-field scattering pattern shown in (b).
}
\end{figure}

The upper results demonstrate that structured nonmagnetic nanoparticles can be used to realize the \textit{ideal MD scattering}. It should be pointed out that structured light can also be used to realize ideal MD scattering as the MD mode in silicon nanoparticle resembles the electromagnetic field distributions for an azimuthally polarized beam \cite{PRL2015,Purcell4}.

In summary, we demonstrate both analytically and numerically that Au(core)/Si(shell) nanoparticles, including spherical and disk geometries, can be used to realized the \textit{ideal MD scattering}, where other multipole modes (including the ED mode) are suppressed, with tunable resonant wavelengths at the near infrared spectrum. These types of nanoparticles provide an ideal platform to study the magnetic light-matter interaction in nanoscale. Based on our results, we expect that functional magnetic light based devices can be realized, such as pure magnetic mirror and ideal directional magnetic scattering. 

Y. Xu and T. Feng thank the financial support from National Natural Science Foundation of China (Grant Nos. 11674130 and 11304047); Natural Science Foundation of Guangdong Province, China (Grant Nos. 2016A030306016, 2014A030313376);

\section{Supplemental Material of 'Ideal magnetic dipole scattering'}% Force line breaks with \\

\author{Tianhua Feng$^{1}$, Yi Xu$^{1,2}$}
\email{e$_$chui@qq.com}
\author{Wei Zhang$^{1}$}
\author{Andrey E. Miroshnichenko$^{3}$}
\email{andrey.miroshnichenko@anu.edu.au}
\affiliation{$^{1}$JNU-IREE Joint Laboratory of Fractal Method and Signal Processing, Department of Electronic Engineering, College of Information Science and Technology, JiNan University, Guangzhou, 510632, China}
\affiliation{$^{2}$Key Laboratory of Optoelectronic Information and Sensing Technologies of Guangdong Higher Education Institutes, Guangzhou, Guangdong 510632, China}
\affiliation{$^{3}$Nonlinear Physics Centre, Research School of Science and Engineering, Australian National University, Canberra ACT 0200, Australia}

\date{\today}% It is always \today, today,

%\pacs{}% PACS, the Physics and Astronomy
\maketitle

%\tableofcontents

\section{The conditions for pure magnetic dipole scattering}
For a spherical silicon nanoparticle (SN), the contributions of spherical electric dipole (ED) and magnetic dipole (MD) to the total scattering can be read as:
\begin{eqnarray}
\label{eq:eq1}
Q_{sca}=\frac{6}{k^2R^2}|a_{1}|^2
\end{eqnarray}
\begin{eqnarray}
\label{eq:eq2}
Q_{sca}=\frac{6}{k^2R^2}|b_{1}|^2
{}\end{eqnarray}
\begin{eqnarray}
\label{eq:eq3}
a_{1}=\frac{[D_{1}(mkR)/m+1/(kR)]\psi_{1}(kR)-\psi_{0}(kR)}{[D_{1}(mkR)/m+1/(kR)]\xi_{1}(kR)-\xi_{0}(kR)}
\end{eqnarray}
\begin{eqnarray}
\label{eq:eq4}
b_{1}=\frac{[mD_{1}(mkR)+1/(kR)]\psi_{1}(kR)-\psi_{0}(kR)}{[mD_{1}(mkR)+1/(kR)]\xi_{1}(kR)-\xi_{0}(kR)}
\end{eqnarray}
where $k$ = 2$\pi/\lambda$ is the wavenumber, $\lambda$ is the wavelength in vacuum, $R$ is the radius of the nanoparticle and $m$ is the ratio between the refractive indices of Si and background medium (air in this paper). $D_{n}(mkR)=\psi_{n}'(mkR)/\psi_{n}(mkR)$, $\psi_{n}(z)$ and $\xi_{n}(z)$ are the Riccati-Bessel functions, $z$ = $kR$ or $mkR$ and $a_{1}$ and $b_{1}$ are the Mie scattering coefficients.  The contributions of high order electric and magnetic multipoles to the total scattering are assumed to be negligible. 

Therefore, the conditions to realize an \textit{ideal magnetic dipole scatter} are the solutions ($\lambda, R$) of this equation set:
\begin{eqnarray}
\label{eq:eq5}
\left\lbrace
\begin{aligned}
(D_{1}(mkR)/m+1/(kR))\psi_{1}(kR)-\psi_{0}(kR)=0\\|b_{1}(\lambda)|_{R}'=0;,\;\;|b_{1}(\lambda)|_{R}''<0\\|a_{n}(\lambda,R)|,|b_{n}(\lambda,R)|=0,n>1
\end{aligned}
\right.
\end{eqnarray}
where the prime means differentiation with respect to the argument in parentheses. In realistic case, the \textit{ideal MD scattering} condition can be slightly released as the solutions of the following equation set:
\begin{eqnarray}
\label{eq:eq6}
\left\lbrace
\begin{aligned}
|(D_{1}(mkR)/m+1/(kR))\psi_{1}(kR)-\psi_{0}(kR)|\ll 1\\|b_{1}(\lambda)|_{R}'=0;,\;\;|b_{1}(\lambda)|_{R}''<0\\|a_{n}(\lambda,R)|,|b_{n}(\lambda,R)|\ll 1,n>1
\end{aligned}
\right.
\end{eqnarray}
Unfortunately, this equation set also has no solution in the visible to near infrared wavelength range.

Alternatively, a core/shell geometry offers additional tunability, where the contributions of spherical ED and MD to the total scattering also can be calculated by Eqs. (1) and (2) with
\begin{eqnarray}
\label{eq:eq7}
a_{1}=\frac{[\widetilde{D}_{1}/m_{2}+1/(kR)]\psi_{1}(kR)-\psi_{0}(kR)}{[\widetilde{D}_{1}/m_{2}+1/(kR)]\xi_{1}(kR)-\xi_{0}(kR)}
\end{eqnarray}
\begin{eqnarray}
\label{eq:eq8}
b_{1}=\frac{[\widetilde{G}_{1}m_{2}+1/(kR)]\psi_{1}(kR)-\psi_{0}(kR)}{[\widetilde{G}_{1}m_{2}+1/(kR)]\xi_{1}(kR)-\xi_{0}(kR)}
\end{eqnarray}
The realization of \textit{ideal magnetic dipole scattering} requires the following equation set has solutions in the optical spectrum:
\begin{eqnarray}
\label{eq:eq9}
\left\lbrace
\begin{aligned}
|(\widetilde{D}_{1}/m_{2}+1/(kR))\psi_{1}(kR)-\psi_{0}(kR)|\ll 1\\|b_{1}(\lambda)|_{r,R}'=0\\|b_{1}(\lambda)|_{r,R}''<0\\|a_{n}(\lambda,r,R)|,|b_{n}(\lambda,r,R)|\ll 1,n>1
\end{aligned}
\right.
\end{eqnarray}
where 
\begin{eqnarray}
\label{eq:eq10}
\widetilde{D}_{1}=\frac{D_{1}(m_{2}kR)-A_{1}\chi_{1}'(m_{2}kR)/\psi_{1}(m_{2}kR)}{1-A_{1}\chi_{1}(m_{2}kR)/\psi_{1}(m_{2}kR)}
\end{eqnarray}
\begin{eqnarray}
\label{eq:eq11}
\widetilde{G}_{1}=\frac{D_{1}(m_{2}kR)-B_{1}\chi_{1}'(m_{2}kR)/\psi_{1}(m_{2}kR)}{1-B_{1}\chi_{1}(m_{2}kR)/\psi_{1}(m_{2}kR)}
\end{eqnarray}
\begin{eqnarray}
\label{eq:eq12}
A_{1}=\frac{\psi_{1}(m_{2}kr)(mD_{1}(m_{1}kr)-D_{1}(m_{2}kr))}{mD_{1}(m_{1}kr)\chi_{1}(m_{2}kr)-\chi_{1}'(m_{2}kr)}
\end{eqnarray}
\begin{eqnarray}
\label{eq:eq13}
B_{1}=\frac{\psi_{1}(m_{2}kr)(D_{1}(m_{1}kr)/m-D_{1}(m_{2}kr))}{D_{1}(m_{1}kr)\chi_{1}(m_{2}kr)/m-\chi_{1}'(m_{2}kr)}
\end{eqnarray}
where $k$ = 2$\pi/\lambda$ wavenumber, $\lambda$ is the wavelength in vacuum, $r$ and $R$ are the inner and outer radius of the core/shell nanoparticles, respectively, which are independent variables of Eq. (9) and $m = m_{2}/m_{1}$ with $m_{1}$ and $m_{2}$ are relative refractive indices of inner and outer mediums. This equation set can be solved numerically. We truncate this equation set use $n$ = 2 here.

\section{Plane wave scattering of Au(core)/silicon(Shell) nanoparticle}
As can be seen from the plane wave scattering results shown in Fig. S1, the ED and MD modes in an Au(core)/silicon(shell) nanoparticle might overlap in spectrum, leading to the unidirectional scattering around 1.34 $ \mu m $.
\begin{figure}[!htb]
\renewcommand\thefigure{S1}
\includegraphics[width=\columnwidth]{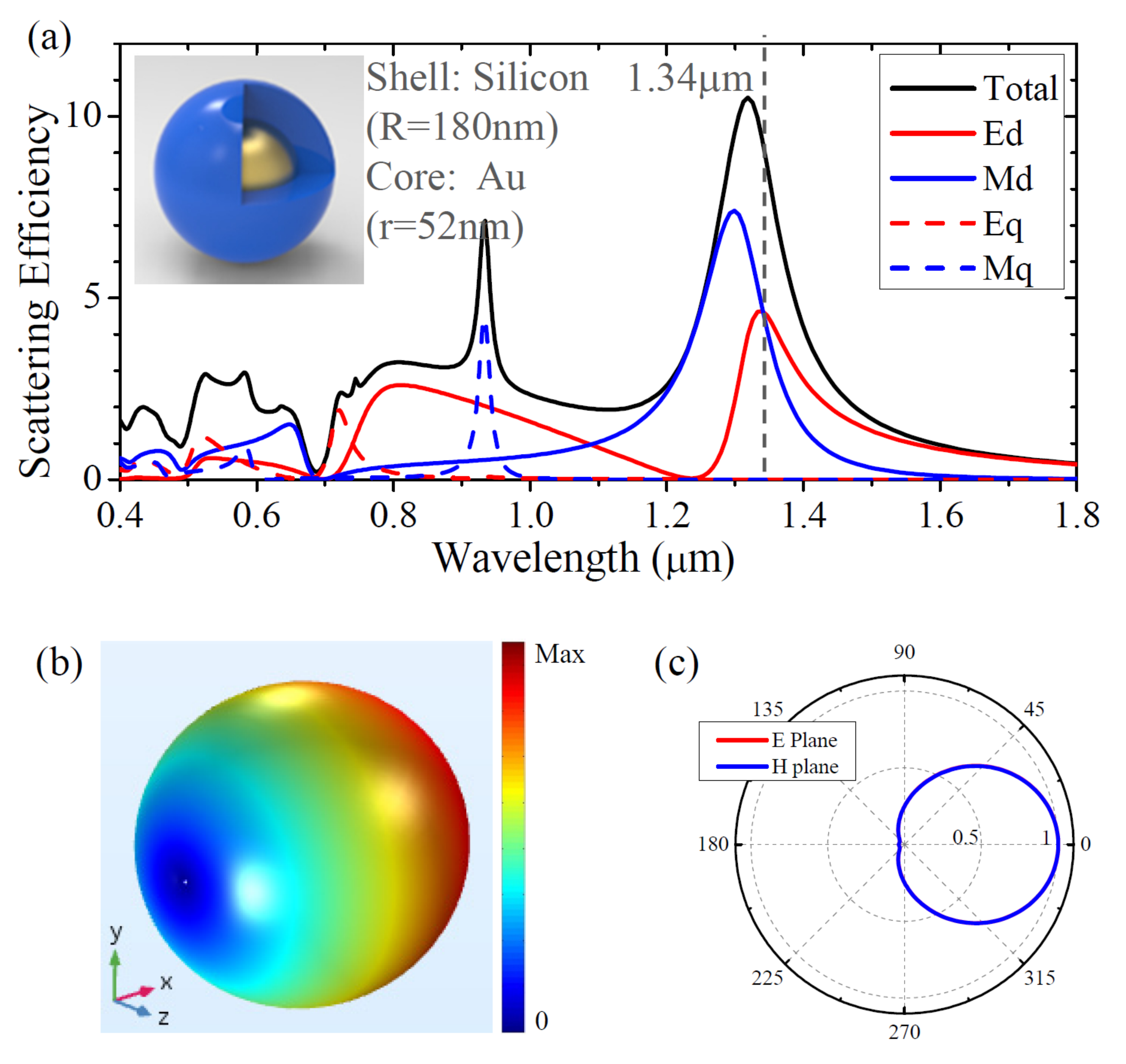}
\caption{\label{fig:fig1} 
(Color online) (a) Scattering efficiency of a Au(core)/silicon(shell) nanoparticle embedded in air with an outer radius $R$ = 180 nm and an inner radius $r$ = 52 nm. The contributions of spherical ED, MD, electric quadrupole (EQ) and magnetic quadrupole (MQ) to the total scattering are also presented. Sketch map shows the geometry of the core/shell nanoparticle. (b) Far field scattering pattern at the wavelength marked by the dashed line in (a). A plane wave incidents along x-axis with its electric polarization along y-axis. (c) Cross section views of the far field scattering pattern shown in (b).
}
\end{figure}

\section{Tunable \textit{ideal MD scattering} in nanospheres}

\textit{Ideal MD scattering} with tunable resonant wavelengths can be realized by changing the sizes of the Au core and silicon shell accordingly. As can be seen from Fig. S2, the resonant frequencies of pure MD mode can be tuned from $\sim$ 1000 nm to $\sim$ 1700 nm. At such resonant regions, the contributions from high order modes are all negligible, validating the \textit{ideal MD scattering} in the near infrared wavelength spectrum.

\begin{figure}[!htb]
\renewcommand\thefigure{S2}
\includegraphics[width=\columnwidth]{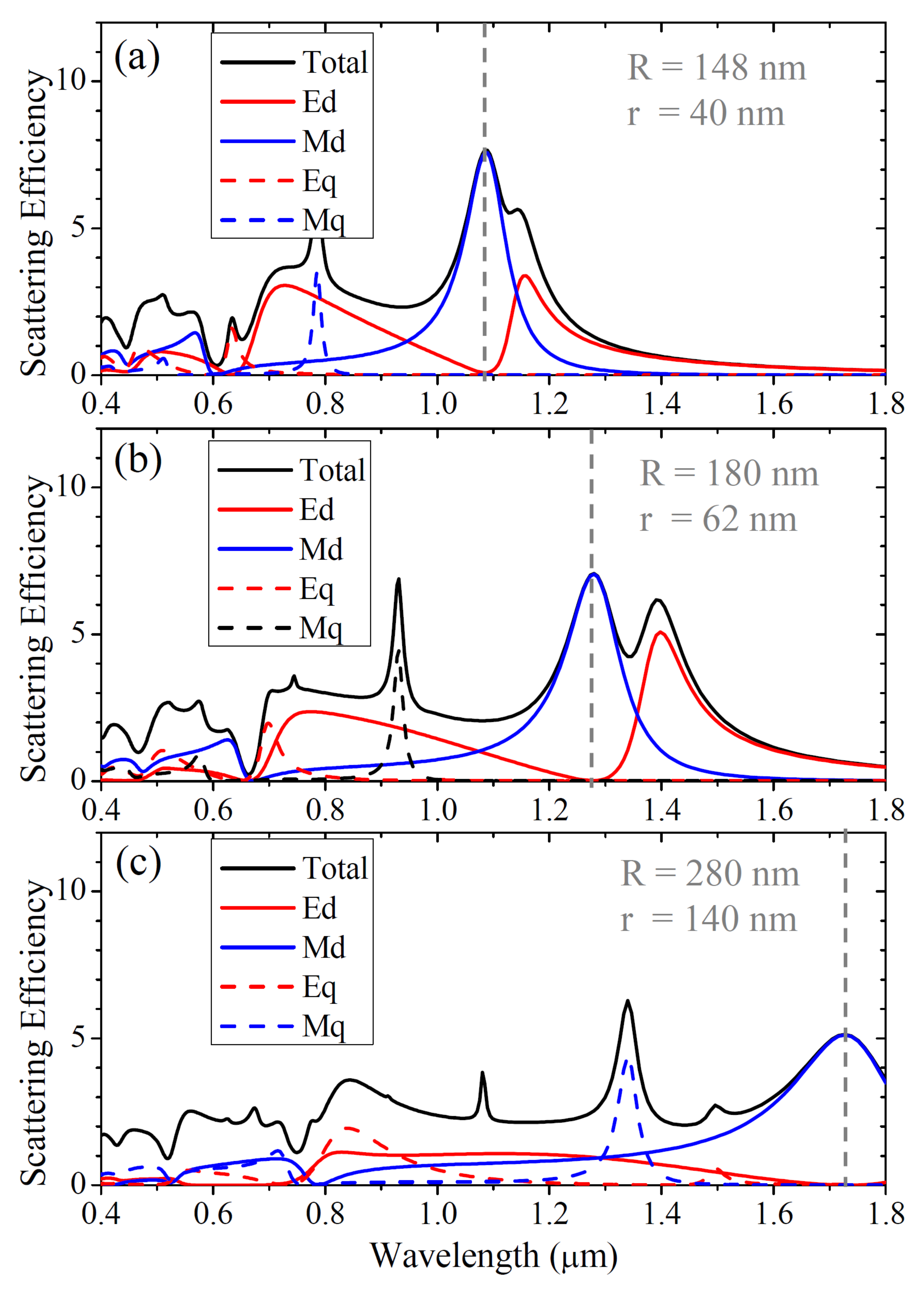}
\caption{\label{fig:fig2} 
(Color online) (a) Scattering efficiency of a Au(core)/silicon(shell) nanoparticle embedded in air with an outer radius $R$ = 148 nm and an inner radius $r$ = 40 nm. The contributions of spherical ED, MD, EQ and MQ to the total scattering are also presented. (b) and  (c) are similar to (a) except for $R$ = 180 nm, $r$ = 62 nm and $R$ = 280 nm, $r$ = 140 nm, respectively. Vertical dashed lines indicate the resonant frequencies where the \textit{ideal MD scattering} appears.
}
\end{figure}

\section{Origin of anapole mode in Au(core)/silicon(shell) nanodisk}
In order to explore the contribution of anapole mode to the \textit{ideal magnetic dipole scattering}, we present the contributions of Cartesian electric and toroidal dipole (TD) modes to the total scattering, as shown in Fig. S3. Similar to the results of spherical nanoparticle presented in Fig. 4, the results in the lossless case indicate that the destructive interference between the Cartesian ED and TD results in zero response of spherical ED in the far-field scattering. The overlap of anapole mode and MD resonance facilitates the realization of \textit{ideal MD scattering} in core/shell nanodisk.
\begin{figure}[!htb]
\renewcommand\thefigure{S3}
\includegraphics[width=\columnwidth]{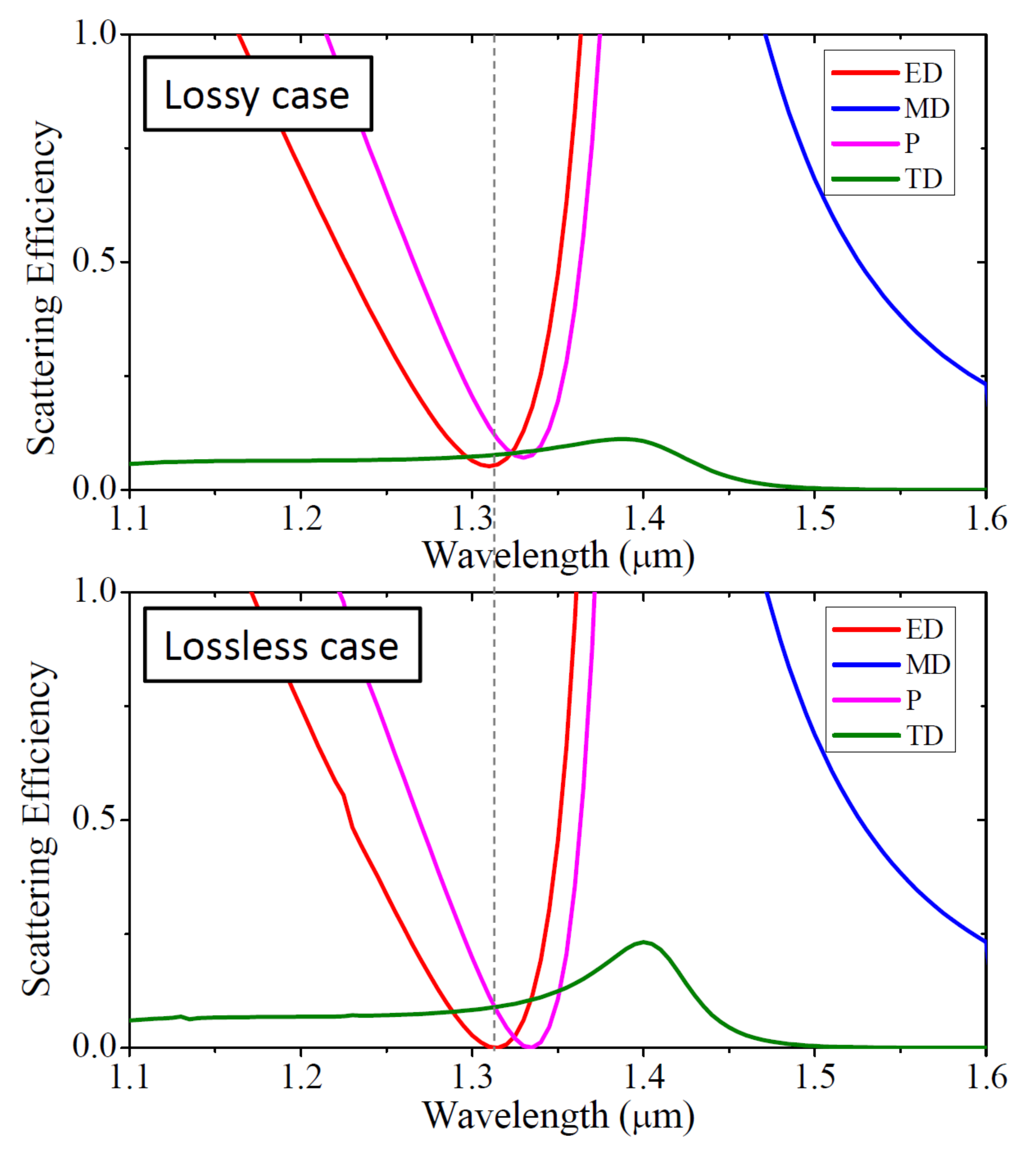}
\caption{\label{fig:fig3} 
(Color online) Spherical ED (red) and MD (blue), Cartesian ED (P, purple) and TD (green) dipole moments contributions to the total scattering of core/shell nanoparticle shown in Fig. 5 for lossy (a) and lossless (b) cases. Vertical dashed line indicates the resonant frequency where the ideal MD scattering appears.
}
\end{figure}

\section{Role of imperfections }
We also consider the effects of structure imperfections on the performance of the core/shell ideal MD scatter based on the nanodisk. As shown in Fig. S4, small deviations of the positions of the Au core have only a minor effect on the ideal MD scattering. Special attention should be paid to the fabrication process of this kind of nanostructure to preserve the exotic property of the ideal MD scatter. 
\begin{figure}[H]
\renewcommand\thefigure{S4}
\includegraphics[width=\columnwidth]{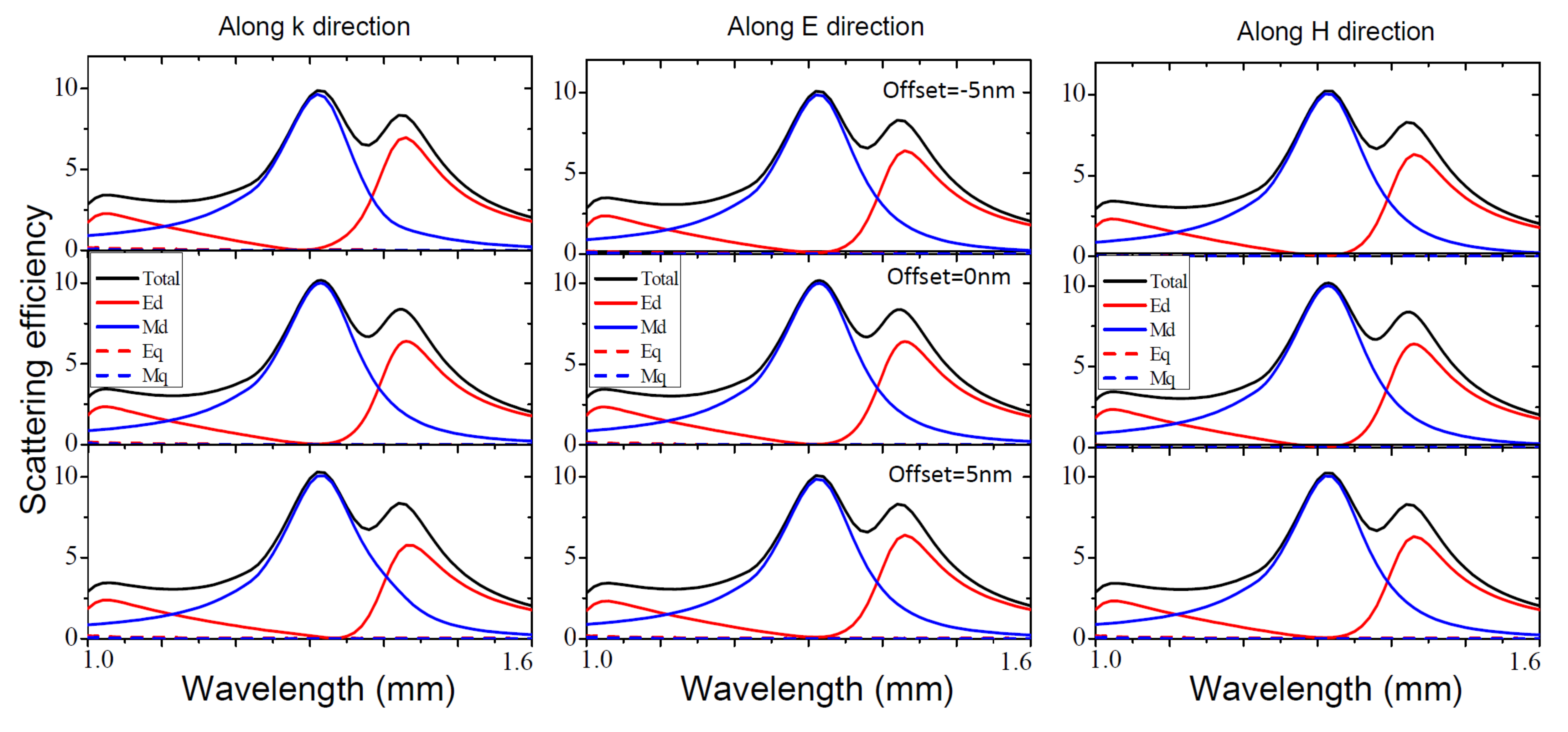}
\caption{\label{fig:fig4} 
(Color online) The effect of 5 nm displacement of the Au core along three directions (x,y,z-axes) on the scattering of the core/shell nanoparticle.
}
\end{figure}

\end{document}